\documentclass[12pt,aps,english]{revtex4}
\usepackage{graphicx}
\usepackage{bm}
\usepackage{dcolumn}

\begin{document}

\title{Looping dynamics of a flexible chain with internal friction at different degrees of compactness}
\author{Nairhita Samanta and Rajarshi Chakrabarti*}
\affiliation{Department of Chemistry, Indian Institute of Technology Bombay, Mumbai, Powai 400076, E-mail: rajarshi@chem.iitb.ac.in}
\date{\today}

\begin{abstract}

Recently single molecule experiments have shown the importance of internal friction in biopolymer dynamics.  Such studies also suggested that the internal friction although independent of solvent viscosity has strong dependence on denaturant concentration. Recent simulations also support such propositions by pointing out weak interactions to be the origin of internal friction in proteins.  Here we made an attempt to investigate how a single polymer chain with internal friction undergoes reconfiguration and looping dynamics in a confining potential which accounts for the presence of the denaturant, by using recently proposed ``Compacted Rouse with internal friction (CRIF)". We also incorporated the effect of hydrodynamics by extending this further to ``Compacted Zimm with internal friction (CZIF)".  All the calculations are carried out within the Wilemski Fixmann (WF) framework. By changing the strength of the confinement we mimicked chains with different degrees of compactness  at different denaturant concentrations.  While compared with experiments our results are found to be in good agreement.
\end{abstract}

\maketitle
\section{Introduction}

 In recent past there have been efforts based on single molecule experiments to elucidate the role of internal friction in protein folding \cite{schulernaturecomm2012, schuler2012}. These single molecule experiments showed internal friction to play a significant role in folding especially when the protein starts with a more compact unfolded state. Among the different types of experiments carried out to study the dynamics of the unfolded proteins, the most common one is a combination of FRET and nsFCS, where two residues $n$ and $m$ of a protein are labelled with fluorescence probes and the fluctuation of the distance between them is observed from the efficiency of energy transfer. An auto-correlation function is then calculated from the distribution, which eventually gives a characteristic relaxation time, also called the reconfiguration time ($\tau_{nm}$) \cite{schulernaturecomm2012}. In another type of fluorescence experiment, two different residues of a protein are again tagged with a fluorescence quencher and donor. The time required for the the donor to be quenched which happens only within a certain distance, gives an estimation of loop formation time ($\tau_{nm,loop}$) in a protein \cite{lapidus2000}. Both the time scales, $\tau_{nm}$ and $\tau_{nm,loop}$ seem to have a nonzero intercept when plotted against the solvent viscosity. This residual reconfiguration or looping time has been then attributed to ``internal friction" which is present within the protein and assumed to be independent of the solvent around. Unfortunately a molecular picture of this internal friction is still lacking but it is believed that dihedral rotations, hydrogen bonding and other weak interactions contribute to internal friction. Recent simulations on proteins support such propositions as well \cite{netzepje, netzjacs, best2014, makarov2014}. This also tells why internal friction is more when the protein is in more compact unfolded state. Though the notion of ``internal friction" is not very old in the physical chemistry community \cite{thirumalaijcp2000, wolynes2001}, it has always been a topic of research among polymer rheologists. For example Rabin and \"{O}ttinger looked at the origin of internal viscosity in a Gaussian chain \cite{rabin1990, schieber}. Following an idea of de Gennes \cite{degennes} they derived an expression for the relaxation time, $\tau_{rel}$ associated with internal viscosity as $\tau_{rel}=R^3/k_BT(\eta_s+\eta_i)$ where, $R=aN^{\nu}$ and $a$, $N$ are the monomer size and chain the length respectively, $\nu$ is the Flory exponent \cite{floryppc, florysmcm, rubinstein}. Therefore in the limit solvent viscosity $\eta_s\rightarrow 0$,  it has a non-zero intercept proportional to the internal viscosity $\eta_i$. This is in the same spirit as that of Khatri and McLeish \cite{khatri2007}, where a modified Rouse model gives a mode relaxation time which is dependent on internal friction, $\tau_p^{RIF}=\tau^R/p^2+\tau_{int}$. Such a model gives a reconfiguration time between any two monomers $n$ and $m$ as  $\tau_{nm}\simeq0.82\tau^R+\tau_{int}$ \cite{chakrabarti2013}, where $\tau^R$ is the rouse time, proportional to the solvent viscosity. Although such a model qualitatively can predict the trend of the variation of reconfiguration time as a function of solvent viscosity and produce a non-zero intercept but cannot account for the changes in ``internal friction" at different degrees of compactness encountered in experiments. It is expected that the protein experiences different level of ``internal friction" at different concentrations of the denaturant as the compactness of it changes. This can be seen from the plots of the reconfiguration time \cite{schuler2012} against the solvent viscosity at different denaturant concentrations where the intercepts gives the time scale due to ``internal friction". The higher the denaturant concentration the smaller the intercept.  In the present study we extend recently proposed models \cite{chakrabarti2013, chakrabarti2014} to include the effect of the compactness of the polymer chain to internal friction by introducing a confining harmonic potential to each monomer of the chain which accounts for the change in chain conformation due to denaturant. At lower denaturant concentration chain is more compact so the confining potential is steeper but as the denaturant concentration increases the confining potential becomes shallower. With this model, the looping dynamics is studied within Wilemski Fixman(WF) framework \cite{wilemski1974} assuming the polymer chain to be Gaussian. Loop formation between any two parts of a bio-polymer is supposedly the primary step of protein folding, DNA cyclization \cite{snigdha2014, metzler2015}. It is worth mentioning that WF formalism seems to work fine here and has been used to calculate the same in presence of hydrodynamic interactions by Chakrabarti \cite{chakrabartiphysica1}  and the effect of viscoelastic solvent \cite{chakrabartiphysica2, bhattacharyya}. The method has also been used in the past extensively \cite{bagchi2001, sebastianjcp, cherayiljcp2004, cherayiljcp20021, cherayiljcp20022, santo} to investigate other aspects of the polymer looping problems.

 The paper is arranged as follows. The details of the the polymer models are given in section $\bf{II}$. Section $\bf{III}$ deals with the methods. Results and discussions are represented in section $\bf{IV}$. Section $\bf{V}$ is the conclusion.

\section{Polymer model}

In the Rouse model, a polymer chain is imagined as a series of brownian particles connected by harmonic springs with no hydrodynamic interactions and excluded volume effect \cite{doibook, kawakatsubook}. If $R_n(t)$ is the position of the $n^{th}$ monomer at time $t$, where $n$ can be varied from $0$ to $N$, then dynamics of Rouse chain with $(N+1)$ monomers is described by the following equation of motion

\begin{equation}
\zeta \frac{\partial{R_{n}(t)}}{\partial{t}}=k\frac{\partial^2{R_{n}(t)}}{\partial{n^2}}+f(n,t)
\label{eq:rouse-model}
\end{equation}

\noindent Where $\zeta$ is the solvent viscosity and $k=\frac{3 k_B T}{b^2}$ is the spring constant with kuhn length $b$. $f(n,t)$ is the random force with moments

\begin{equation}
   \left<f(n,t)\right>=0,
   \left<f_{\alpha}(n,t_1)f_{\beta}(m,t_2)\right>=2 \zeta k_B T \delta_{\alpha\beta}\delta(n-m)\delta(t_1-t_2)
\label{eq:random-force}
\end{equation}

Standard procedure to treat such a system is to describe the dynamics in the form of normal modes $(X_p)$

\begin{equation}
\zeta_p^R \frac{d{X_{p}(t)}}{d{t}}=-k_p^R X_{p}(t)+f_p(t)
\label{eq:rouse-mode}
\end{equation}

\noindent Here, $\zeta_p^R=2 N \zeta$ for $p>0$ and $\zeta_0^R=N \zeta$. The relaxation time of $p^{th}$ mode is $\tau_p^R=\frac{\zeta_p^R}{k_{p}^R}=\frac{\tau^R}{p^2}$ and $k_p^R=\frac{6 \pi^2 k_B T p^2}{N b^2}$. The slowest relaxation time $\tau^R=\frac{N^2 \zeta b^2}{3 \pi^2 k_B T}$ is called Rouse time.

Rouse model does not take into account the effects of hydrodynamic interactions. The simplest possible  model which takes care of it, is due to Zimm \cite{doibook, kawakatsubook}. It is possible to show that in $\theta$ condition  under pre-averaged hydrodynamic interaction, the Zimm chain can be described by the same Eq. (\ref{eq:rouse-mode}) but with  a different scaling of friction co-efficient with the mode number $\zeta_p^Z =\zeta\left(\frac{\pi N p}{3}\right)^{1/2}$ where, $k_p$ remains the same i.e. $k_p^Z=k_p^R=k_p^{R(Z)}$ \cite{chakrabartiphysica1}. The relaxation time of $p^{th}$ mode is $\tau_p^Z=\frac{\zeta_p^Z}{k_p^Z}=\frac{\tau^Z}{p^{3/2}}$ and $\tau^Z=\frac{N^{3/2} \zeta b^2}{6\sqrt{3} \pi^{3/2} k_B T}$ , is the slowest relaxation time or Zimm time.

Recently in polymer rheology community internal friction has been introduced within Rouse description using a dashpot between the neighbouring monomers. This is commonly referred to as Rouse with internal friction (RIF) which has the following euation of motion

\begin{equation}
\zeta \frac{\partial{R_{n}(t)}}{\partial{t}}=\left(k+\zeta_{int}\frac{\partial}{\partial{t}}\right)\frac{\partial^2{R_{n}(t)}}{\partial{n^2}}+f(n,t)
\label{eq:rif-model}
\end{equation}

 \noindent Khatri and McLeish \cite{khatri2007} showed that above model can also be treated with normal mode description as is done with Rouse chain. Similarly a Zimm chain with internal friction (ZIF) can be constructed. In both the models RIF and ZIF, internal friction appears as an additive constant to the relaxation times of each normal modes $(p>0)$ \cite{makarov2013,chakrabarti2013}. The new relaxation time is  $\tau^{R(Z)IF}_p=\tau_p + \tau_{int}$ where, $\tau_{int}=\zeta_{int}/k$. This is because of the redefined $\zeta_p^{R(Z)IF}=\zeta_p^{R(Z)}+2\pi^2p^2\zeta_{int}/N, p>0$ where $k_p^{R(Z)}$ remains unchanged, i.e. $k_p^{R(Z)IF}=k_p^{R(Z)}$.

In RIF and ZIF there is no scope to introduce the effect of denaturant concentration to the chain conformation. It is obvious that the more is the denaturant concentration, the less compact the chain is and the less is the internal friction. To incorporate such effect, one possibility is to put all the monomers in a confining potential. This becomes analytically trackable when the confinement is harmonic. Thus the restoring force acting on the $n^{th}$ monomer would be $-\frac{\partial}{\partial{R_n}}(\frac{k_c}{2}(R_n-0)^2)$ where, $k_c$ is the spring constant.  Similar models of polymers in confined potential has been used in other contexts too \cite{sebastianjacs, sebastianpre1, klspre, sebastianpre2, alokpre, debnathpre}. This model is named as Compacted Rouse with internal friction (CRIF). We would also like to mention that essentially the same model has recently been used to fit the simulated results \cite{makarov2014}.

\begin{equation}
\zeta \frac{\partial{R_{n}(t)}}{\partial{t}}=\left(k+\zeta_{int}\frac{\partial}{\partial{t}}\right)\frac{\partial^2{R_{n}(t)}}{\partial{n^2}}-k_c R_{n}(t)+f(n,t)
\label{eq:crif-model}
\end{equation}

\noindent Similarly ZIF can be extended to Compacted Zimm with internal friction (CZIF). Now, $\tau_p^{CRIF}$ as well as $\tau_p^{CZIF}$ are mode dependent because $k_p^{CR(Z)IF}=k_{p}^{R(Z)}+2 N k_c$, but the friction coefficients associated with normal modes remain unchanged i.e. $\zeta_p^{CRIF}=\zeta_p^{RIF}$ and $\zeta_p^{CZIF}=\zeta_p^{ZIF}$. As a consequence, the relaxation time for $p^{th}$ mode of this Rouse chain is, $\tau_p^{CRIF}=\frac{\zeta_p^R}{k_p^{CIF}}+\frac{\zeta_{int}}{k+k_c N^2/p^2 \pi^2}$ and for this Zimm chain in $\theta$ solvent under pre-averaged hydrodynamic interactions is $\tau_p^{CZIF}=\frac{\zeta_p^{Z}}{k_p^{CIF}}+\frac{\zeta_{int}}{k+k_c N^2/p^2 \pi^2}$. The time scale for the internal friction of CRIF as well as CZIF become identical $\tau_{int}^{CR(Z)IF}=\frac{\zeta_{int}}{k+k_c N^2/p^2 \pi^2}$. Depending on the value of $k_c$, mode number $p$ and the chain length two extreme situations can arise. For example if $k_c$ is small, chain is short (small $N$) and $p>>1$, $k+k_c N^2/p^2 \pi^2$ can be approximated as $k$ and then $\tau_{int}^{CR(Z)IF}=\tau_{int}^{RIF}$. Same is true for Zimm chain. This is expected as this corresponds to a very weak confinement of the chain where higher modes rarely contribute. On the other hand, if $k_c$ is large, then $k+k_c N^2/p^2 \pi^2\simeq k_c N^2/p^2\pi^2$ and $\tau_{int}^{CR(Z)IF}\simeq\frac{\zeta_{int}p^2\pi^2}{k_cN^2}$. Therefore in such a situation $k_c$ controls the dynamics of the polymer and higher modes contribute more. All the parameters for Rouse and Zimm chain are depicted in Table \ref{tablerouse} and \ref{tablezimm} respectively.

\begin{table}[tbp]
\begin{tabular}{|c||c|c|c|}
\multicolumn{1}{r}{}
 &  \multicolumn{3}{c}{No Hydrodynamics} \\
\hline Parameter & Rouse & RIF & CRIF   \\ \hline
$\zeta_p$ & $\zeta_p^R=2N\zeta$ & $\zeta_p^{RIF}=\zeta_p^R+2\pi^2p^2\zeta_{int}/N$ & $\zeta_p^{CRIF}=\zeta_p^R+2\pi^2p^2\zeta_{int}/N$  \\ \hline
$k_p$ & $k_p^R=\frac{6 \pi^2 k_B T p^2}{N b^2}$ & $k_p^{RIF}=k_p$ & $k_p^{CRIF}=2Nk_c+k_p$ \\
\hline
$\tau_p$ & $\tau_p^R=\frac{\zeta_p^R}{k_p}=\frac{\tau_R}{p^2}$ & $\tau_p^{RIF}=\frac{\zeta_p^R}{k_p}+\tau_{int}$ & $\tau_p^{CRIF}=\frac{\zeta_p^R}{k_p^{CRIF}}+\tau_{int}^{CRIF}$ \\
\hline
$\tau_{int}$ & $0$ & $\tau_{int}^{RIF}=\zeta_{int}/k$ & $\tau_{int}^{CRIF}=\frac{\zeta_{int}}{k+k_c N^2/p^2 \pi^2}$\\ \hline
\end{tabular}
\vspace{0.5 in}
\caption{List of parameters for Rouse, RIF and CRIF}
\label{tablerouse}
\end{table}

\begin{table}[tbp]
\begin{tabular}{|c||c|c|c|}
\multicolumn{1}{r}{}
 & \multicolumn{3}{c}{Hydrodynamics} \\
\hline  Parameter & Zimm & ZIF & CZIF  \\ \hline
$\zeta_p$  & $\zeta_p^Z =\zeta\left(\frac{\pi N p}{3}\right)^{1/2}$ & $\zeta_p^{ZIF} =\zeta_p^Z+2\pi^2p^2\zeta_{int}/N$ & $\zeta_p^{CZIF} =\zeta_p^Z+2\pi^2p^2\zeta_{int}/N$ \\ \hline
$k_p$ & $k_p^Z=\frac{6 \pi^2 k_B T p^2}{N b^2}$ & $k_p^{ZIF}=k_p$ & $k_p^{CZIF}=2Nk_c+k_p$ \\ \hline
$\tau_p$ & $\tau_p^Z=\frac{\zeta_p^Z}{k_p}=\frac{\tau_Z}{p^2}$ & $\tau_p^{ZIF}=\frac{\zeta_p^Z}{k_p}+\tau_{int}$ & $\tau_p^{CZIF}=\frac{\zeta_p^Z}{k_p^{CZIF}}+\tau_{int}^{CZIF}$ \\
\hline
$\tau_{int}$ & $0$ & $\tau_{int}^{ZIF}=\zeta_{int}/k$ & $\tau_{int}^{CZIF}=\frac{\zeta_{int}}{k+k_c N^2/p^2 \pi^2}$ \\ \hline
 \end{tabular}
\vspace{0.5 in}
\caption{List of parameters for Zimm, ZIF and CZIF}
\label{tablezimm}
\end{table}

\section{Calculation methods}
\subsection{Reconfiguration time and mean square displacement}

The time correlation function of normal modes for Rouse as well as Zimm chain is given by the following equation \cite{doibook}

\begin{equation}
\left<X_{p\alpha}(0)X_{q\beta}(t)\right>=\frac{k_B T}{k_p}\delta_{pq}\delta_{\alpha\beta}\exp\left(-t/\tau_p\right)
\label{eq:xpcorr}
\end{equation}

\noindent One needs to only choose the correct $\tau_p$ to find the correlation for Rouse or RIF and Zimm or ZIF \cite{chakrabarti2014}.

When normal mode correlations of CRIF or CZIF are constructed the basic structure of correlation functions remain the same. Only $k_p$ and  $\tau_p$ need to be replaced by $k_p^{CR(Z)IF}$ and $\tau_p^{CR(Z)IF}$ respectively.

\begin{equation}
\left<X_{p\alpha}(0)X_{q\beta}(t)\right>=\frac{k_B T}{k_p^{CR(Z)IF}}\delta_{pq}\delta_{\alpha\beta}\exp\left(-t/\tau_p^{CR(Z)IF}\right)
\label{eq:cifcorr}
\end{equation}

\noindent Therefore Eq. (\ref{eq:cifcorr}) is a generalized expression. From now on, all the expressions are defined for compacted polymers with internal friction to make the discussion more extensive, from which one can easily get back to RIF and ZIF by taking the limit $k_c\rightarrow0$. Usual Rouse and Zimm models can also be recovered by putting $k_c=0$ and $\zeta_{int}=0$

The distance between $n^{th}$ and $m^{th}$ monomers is  $R_{nm}(t)=2 \sum\limits_{p=1}^\infty X_p(t) [cos(\frac{p \pi n}{N}) - cos(\frac{p \pi m}{N})]$, where $R_{n}(t)={X_0} + 2 \sum\limits_{p=1}^\infty X_p(t)cos(\frac{p \pi n}{N})$ and has the following time correlation function

\begin{equation}
{\phi}_{nm}(t)=\left<{R}_{nm}(0).{R}_{nm}(t)\right>
=4 \sum\limits_{p=1}^\infty \frac{3 k_B T}{k_p^{CR(Z)IF}}[cos(\frac{p \pi n}{N}) - cos(\frac{p \pi m}{N})]^2 exp(-t/ \tau_p^{CR(Z)IF})
\label{eq:corr}
\end{equation}

\noindent Reconfiguration time $\tau_{nm}$ is calculated by taking an time integration of $\tilde\phi_{nm}(t)$ \cite{makarov2010, chakrabarti2013, chakrabarti2014}.

\begin{equation}
\tau_{nm}=\int\limits_{0}^\infty dt \tilde{\phi}_{nm}(t)
\label{eq:recon}
\end{equation}

 \noindent Where, $\tilde\phi_{nm}(t)$ is the normalized correlation function which is defined as

\begin{equation}
\tilde\phi_{nm}(t)=\frac{{\phi}_{nm}(t)}{{\phi}_{nm}(0)}
\label{eq:phit}
\end{equation}

In the limit when $k_c$ is high, $\tau_{nm}=N^2\frac{\sum\limits_{p=1}^\infty \frac{1}{\zeta+\zeta_{int}p^2\pi^2/N^2}[cos(\frac{p \pi n}{N}) - cos(\frac{p \pi m}{N})]^2 }{\sum\limits_{p=1}^\infty \frac{1}{k_c}[cos(\frac{p \pi n}{N}) - cos(\frac{p \pi m}{N})]^2}$, which is independent of $k$ as expected and dynamics is controlled by $k_c$.

Similarly, MSD of $\mathbf{R}_{nm}(t)$ has the following closed form expression,

\begin{equation}
\left<\left({R}_{nm}(t)-{R}_{nm}(0)\right)^2\right>=4 \sum\limits_{p=1}^\infty \frac{6 k_B T}{k_p^{CR(Z)IF}}\left(cos(\frac{p \pi n}{N}) - cos(\frac{p \pi m}{N})\right)^2 \left(1-\ exp(-t/\tau_p^{CR(Z)IF})\right)
\label{eq:msd}
\end{equation}

\subsection{Looping time}

A well known procedure to calculate the looping time is due to  Wilemski-Fixmann (WF) \cite{wilemski1974}. Although the method was initially posed for end-to-end loop formation, but can be generalized for looping between any two monomers of the polymer chain \cite{chakrabarti2014, chakrabartiphysica2}, as long as the chain is Gaussian. So, this method is applicable to CRIF and CZIF as well. Looping time within WF framework is given by following expression,

\begin{equation}
\tau_{nm, loop}=\int_0^{\infty} dt \left( \frac{C_{nm}(t)}{C_{nm}(\infty)}-1\right)
\label{eq:looping-time-mn}
\end{equation}

\noindent Where, $C_{mn}(t)$ is the sink-sink correlation function between the $n^{th}$ and $m^{th}$ monomer

\begin{equation}
C_{nm}(t)=\int dR_{nm} \int dR_{nm,0} S(R_{nm})G(R_{nm},t | R_{nm,0},0)S(R_{nm, 0})P(R_{nm,0})
\label{eq:sink-sink-nm}
\end{equation}

\noindent $G(R_{nm},t | R_{nm,0}, 0)$ is the conditional probability that the distance between $n^{th}$ and $m^{th}$ monomers of the chain, is $R_{nm,0}$ at time $t=0$, and $R_{nm}$ at time $t$. The chain is assumed to be in equilibrium at $t=0$ with distribution, $P(R_{nm,0})=\left(\frac{3}{2\pi \left<R_{nm}^2\right>_{eq}}\right)^{3/2}exp\left[-\frac{3 R_{nm,0}^2}{2 \left<R_{nm}^2\right>_{eq}}\right]$. Where $S(R_{nm})$ is the sink function having infinite strength which depends only in the separation between the two monomers which basically takes care of the loop formation between two monomers. This delta function sink is used in many cases to calculate the survival probability of a brownian particle \cite{bagchibook, szabo1984, sebastianpra1992, debnath2006}.

\noindent The conditional probability is given by,

\begin{equation}
G(R_{nm},t | R_{nm,0},0)=\left(\frac{3}{2\pi \left<R_{nm}^2\right>_{eq}}\right)^{3/2}\left(\frac{1}{(1-\tilde\phi_{nm}^2(t))^{3/2}}\right) exp\left[-\frac{3({R}_{nm}-\tilde\phi_{nm}(t)R_{nm,0})^2}{2\left<R_{nm}^2\right>_{eq}(1-\tilde\phi_{nm}^2(t))}\right]
\label{eq:green-nm}
\end{equation}

\noindent Choosing the sink function $S(R_{mn})$ to be a delta function the expression for looping time becomes

\begin{equation}
\tau_{nm, loop}=\int_0^{\infty} dt \left( \frac{exp[-2 \chi_0 \tilde\phi_{nm}^2(t)/(1-\tilde\phi_{nm}^2(t))]sinh[(2\chi_0\tilde\phi_{nm}(t))/(1-\tilde\phi_{nm}^2(t))]}{(2 \chi_0 \tilde\phi_{nm}(t))\sqrt{1-\tilde\phi_{nm}^2(t)}}-1\right)
\label{eq:taudeltamn}
\end{equation}

\noindent Where,

\begin{equation}
\chi_0=\frac{3 a^2}{2 \left<R_{nm}^2\right>_{eq}}
\label{chi}
\end{equation}

\noindent For CRIF and CZIF $\left<R_{N0}^2\right>_{eq}$ has the following analytical expression

\begin{equation}
\left<R_{N0}^2\right>_{eq}=\frac{2 b \sqrt{3 k_B T}}{\sqrt{k_c}} tanh[\frac{N b \sqrt{k_c}}{2 \sqrt{3 k_B T}}]
\label{req}
\end{equation}

\noindent In the limit, $k_c\rightarrow 0$,

\begin{equation}
\left<R_{N0}^2\right>_{eq,k_c\rightarrow 0}\simeq\frac{2 b \sqrt{3 k_B T}}{\sqrt{k_c}}\frac{N b \sqrt{k_c}}{2 \sqrt{3 k_B T}}=Nb^2
\label{req1}
\end{equation}

Obviously $\left<R_{N0}^2\right>_{eq,k_c\rightarrow 0}>\left<R_{N0}^2\right>_{eq}$ as confining potential makes the polymer more compact and as a consequence the effective kuhn length becomes shorter. Thus the equilibrium distribution function of CRIF and CZIF have narrower widths but remain Gaussian \cite{schuler2012}. Similar situation of renormalization of the kuhn length arises during swelling of a polymer chain \cite{dua2005}, where the kuhn length gets longer.

\section{Results and discussions}

\begin{figure}
   \centering
    \includegraphics[width=0.9\textwidth]{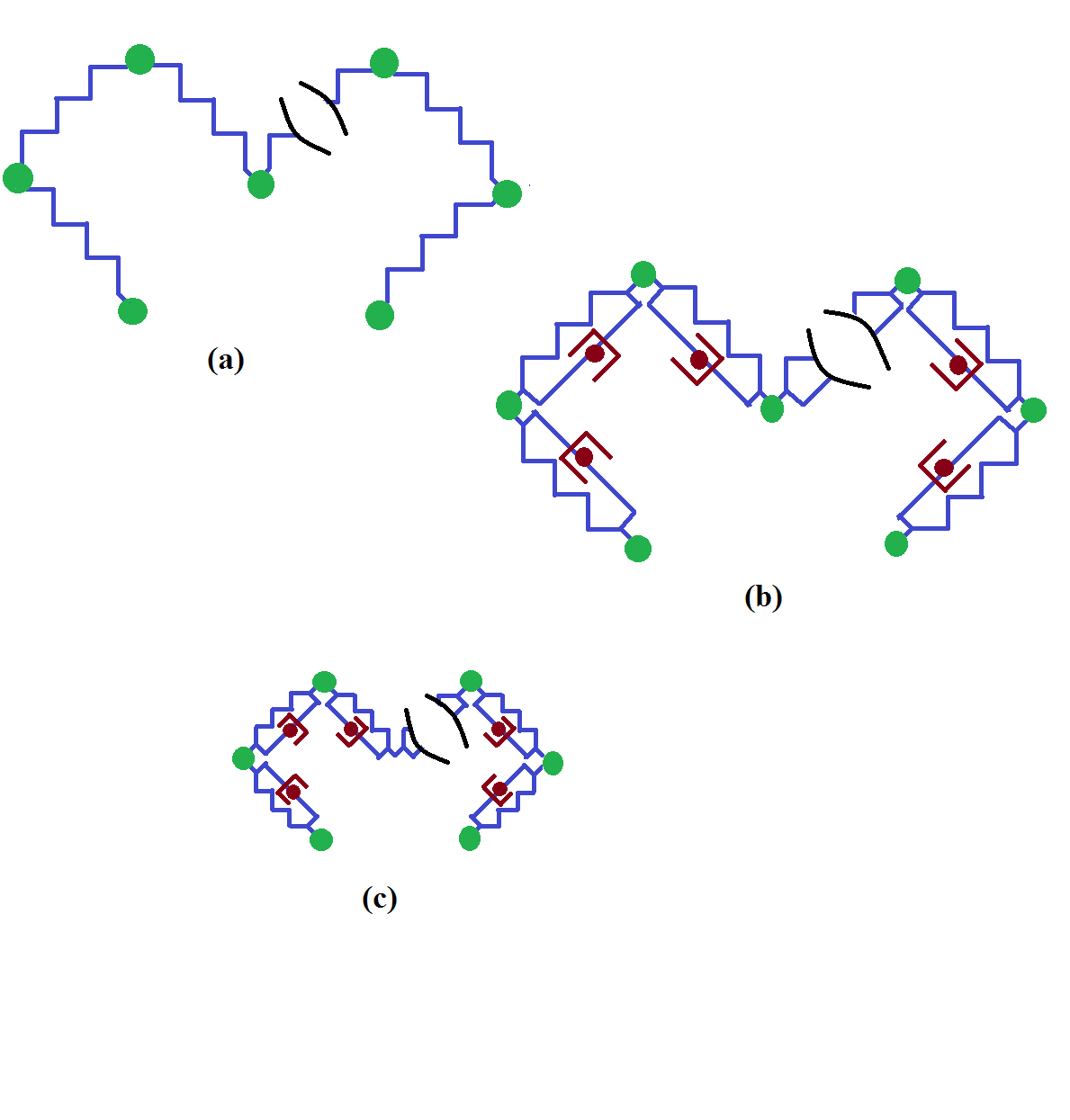}
    \caption{Polymer model (a) Rouse, (b) RIF, (c) CRIF.}
     \label{fig:aa}
\end{figure}

 Schuler and his group studied the dynamics of cold shock protein (Csp) using FRET and FCS. From the auto-correlation function of the distance separating any two arbitrary residues they determined intramolecular diffusion coefficients \cite{schuler2008, schuler2012}. In our analysis all the parameters have been chosen in accordance with these experiments. Thus the  polymer has $67$ monomers $(N + 1 = 67)$, with kuhn length $b = 3.8\times10^{-10}$. In consistance with the viscosity of water, the solvent friction $\zeta = 9.42\times10^{-12} kgs^{-1}$ at temperature $300K$. The force constant of the springs connecting monomers to each other is calculated from the relation $k=3 k_{B}T/b^2$. Without a clear molecular picture of the internal friction, we have to invoke an ansatz which is $\zeta_{int}=\zeta_{int,0}f(n_b)$ \cite{chakrabarti2014}, where $n_b$ is the number non-adjacent monomers contributing to the internal friction and $f(n_b)=(c_0+c_1n_b+c_2n_b^2+....)$, where $\zeta_{int,0}$ is the internal friction due to adjacent neighbour interactions as shown for RIF in Fig. (\ref{fig:aa}) which arises if $c_0=1$ and $c_i=0$ for $i>0$, $n_b=0$. As the solvent quality drops, the polymer shrinks bringing more and more number of monomers together. Hence, $n_b$ and $c_i$ increases so as $f(n_b)$. Taking this relation one step further, we associate $k_c$ with $\zeta_{int}$ with another ansatz $k_c=\tilde{k}_c+A\frac{\zeta_{int,0}}{\tau_{int}}(c_0+c_1n_b+c_2n_b^2+....)$ where, A is typically in the order of $\pi^2/N^2$. Now $k_c$ has two parts $\tilde{k}_c$ and the second part which is proportional to $\zeta_{int}$. This is obvious since the more compact the polymer is, the higher is the internal friction. Again in the limit $c_i=0$ for $i>0$ and $n_b=0$ $k_{c,0}=\tilde{k}_c+A\frac{\zeta_{int,0}}{\tau_{int}}$. $\tilde{k}_c$ accounts for the situation when the chain experiences a harmonic confinement but does not have internal friction. We have considered $\tilde{k}_c$ to be zero in our calculations for the sake of simplicity.

\begin{figure}
  \centering
    \includegraphics[width=0.8\textwidth]{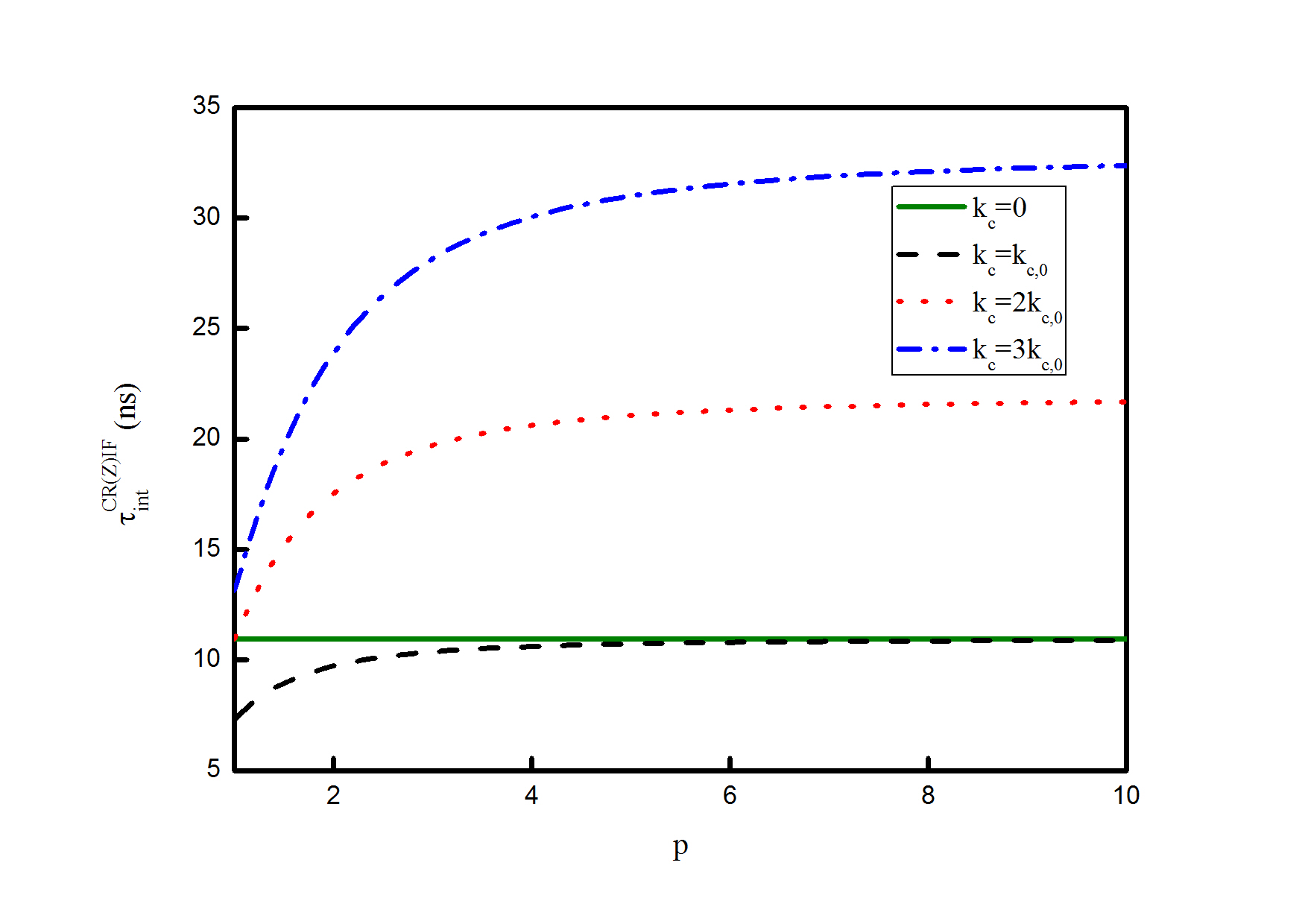}
     \caption{Plot of $\tau_{int}^{CR(Z)IF}$ vs $p$. See text for the values of the parameters used.}
     \label{fig:bb}
   \end{figure}

$\tau_{int}^{CRIF}$ and $\tau_{int}^{CZIF}$ are identical and referred to as $\tau_{int}^{CR(Z)IF}$ in general. Therefore in Fig. (\ref{fig:bb}) the time scale due to the internal friction of each normal mode or $\tau_{int}^{CR(Z)IF}$ is plotted against normal mode number $p$ at three different values of $k_c$ namely $k_{c,0}$, $2k_{c,0}$, $3k_{c,0}$.  As expected it saturates at higher modes which shows that higher modes get rarely affected by the confinement and to the higher modes $\tau_{int}^{R(Z)IF}$ adds as an additive constant as mentioned earlier. It can also be seen from the graph $\tau_{int}^{CR(Z)IF}$ is completely independent of the mode numbers when $k_c$ is zero since in the absence of $k_c$, $\tau_{int}^{CR(Z)IF}$ is nothing but $\tau_{int}^{R(Z)IF}$. With increasing $k_c$, $\zeta_{int}$ also increases which is why $\tau_{int}^{CR(Z)IF}$ becomes larger.

\begin{figure}
  \centering
    \includegraphics[width=0.8\textwidth]{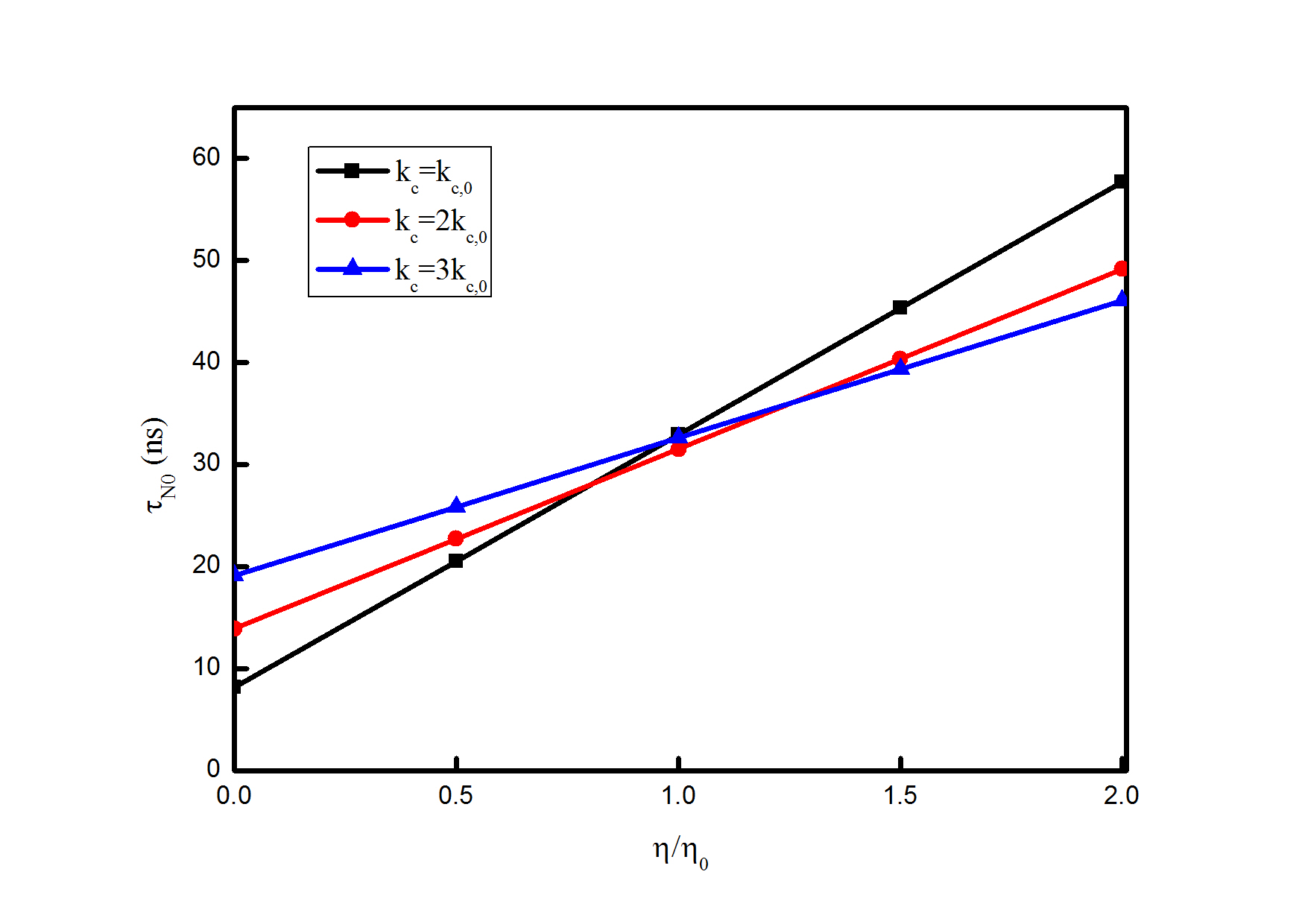}
    \caption{Plot of end-to-end Reconfiguration time vs $\eta/\eta_0$ for CRIF. See text for the values of the parameters used.}
    \label{fig:cc}
\end{figure}

\begin{figure}
   \centering
    \includegraphics[width=0.8\textwidth]{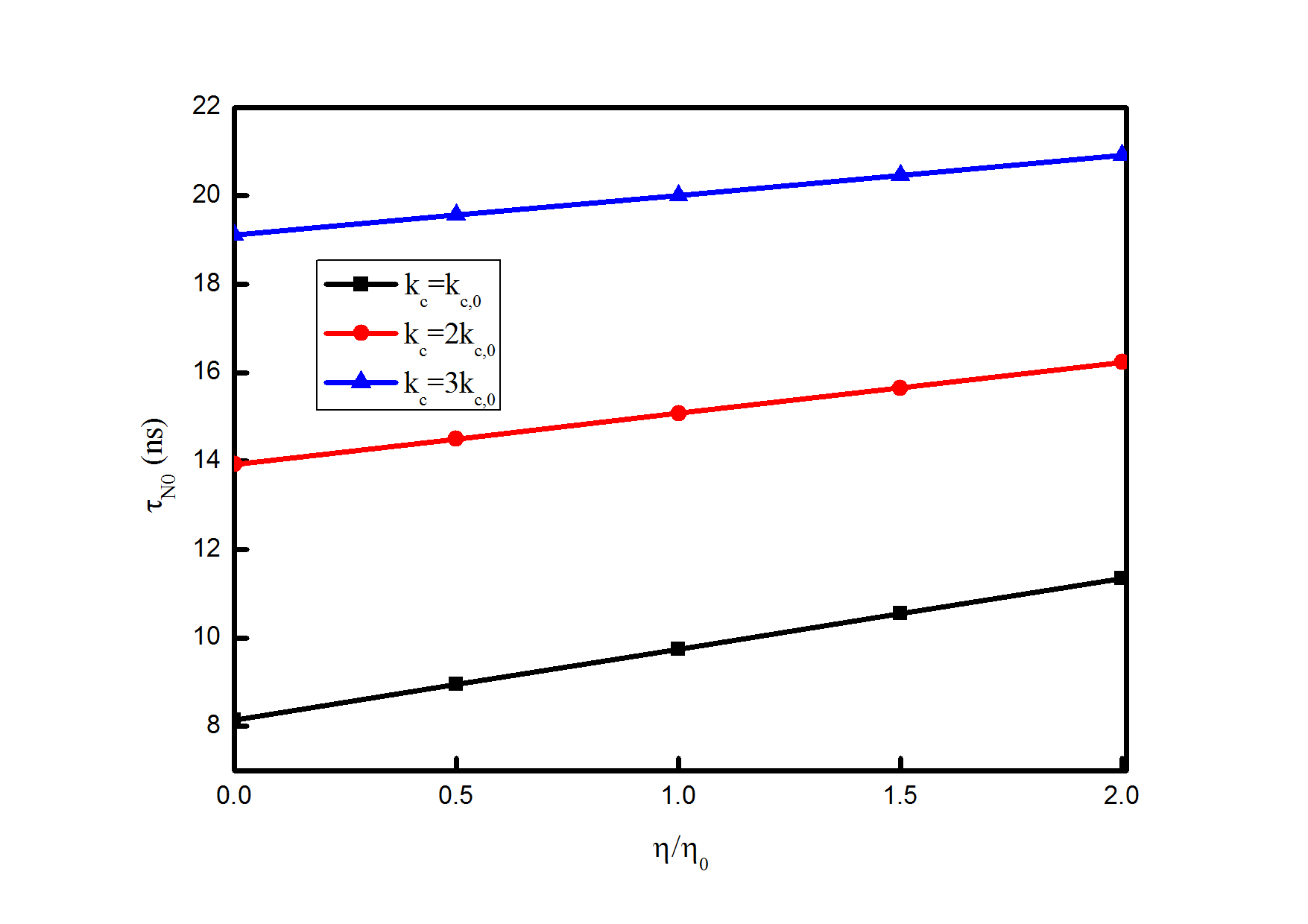}
    \caption{Plot of end-to-end Reconfiguration time vs $\eta/\eta_0$ for CZIF. See text for the values of the parameters used.}
    \label{fig:dd}
\end{figure}

The generalized expression for the reconfiguration time between two monomers $m$ and $n$ is given in Eq. (\ref{eq:recon}), with $m=0$ and $n=N$ it becomes the end-to-end reconfiguration time. End-to-end reconfiguration time is computed at different solvent viscosities $(\eta)$ with three different values of $k_c$ for CRIF. Effects of hydrodynamics have also been taken care of in CZIF. To compare with experiments, reconfiguration times are plotted against $\eta/\eta_0$ in Fig. (\ref{fig:cc}) and Fig. (\ref{fig:dd}) where, $\eta$ is solvent viscosity and $\eta_0$ is viscosity of water. In experiments solvent viscosity is generally varied by adding viscogenic agents such as glycerol to the solution, where as introduction of denaturant such as GdmCl makes a polymer less compact. When plots of $\tau_{N0}$ vs $\eta/\eta_0$ are extrapolated to zero viscosity, the intercepts give the value of internal friction which increases with increasing value of $k_c$ and the slopes give the dependence of reconfiguration time on solvent viscosity. While compared with experiments \cite{schulernaturecomm2012, schuler2012}, the results are consistent with experiments where the intercepts decreases as the denaturant concentration increases (low $k_c$). Typically the time scale due to internal friction calculated from the intercepts are in the range of $10-50$ ns, which are also in good agreement with the experiments. Similar is the case with Zimm chain.

\begin{figure}
  \centering
    \includegraphics[width=0.8\textwidth]{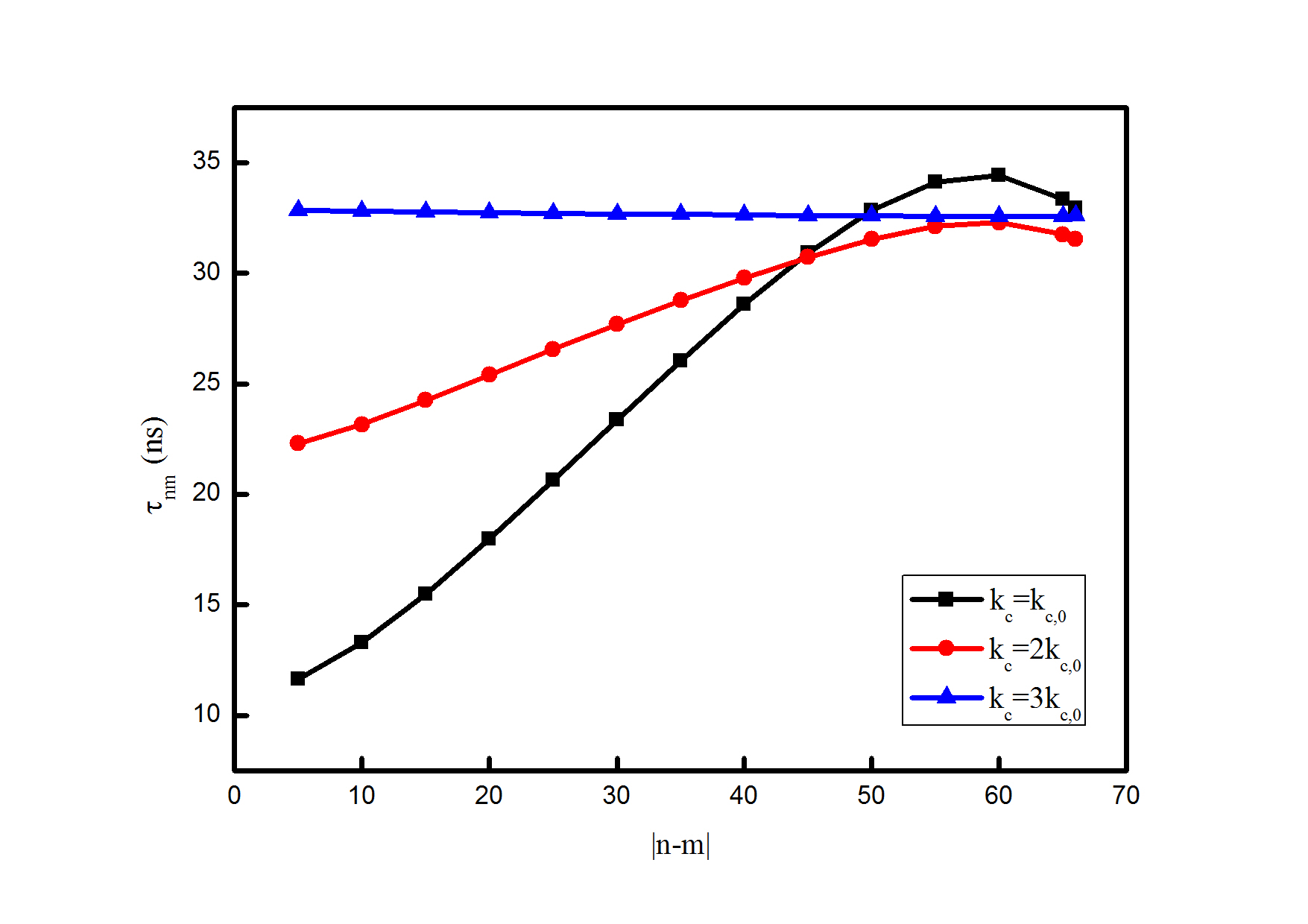}
      \caption{Plot of Reconfiguration time vs $|n-m|$ for CRIF. See text for the values of the parameters used.}
      \label{fig:ee}
\end{figure}

\begin{figure}
  \centering
    \includegraphics[width=0.8\textwidth]{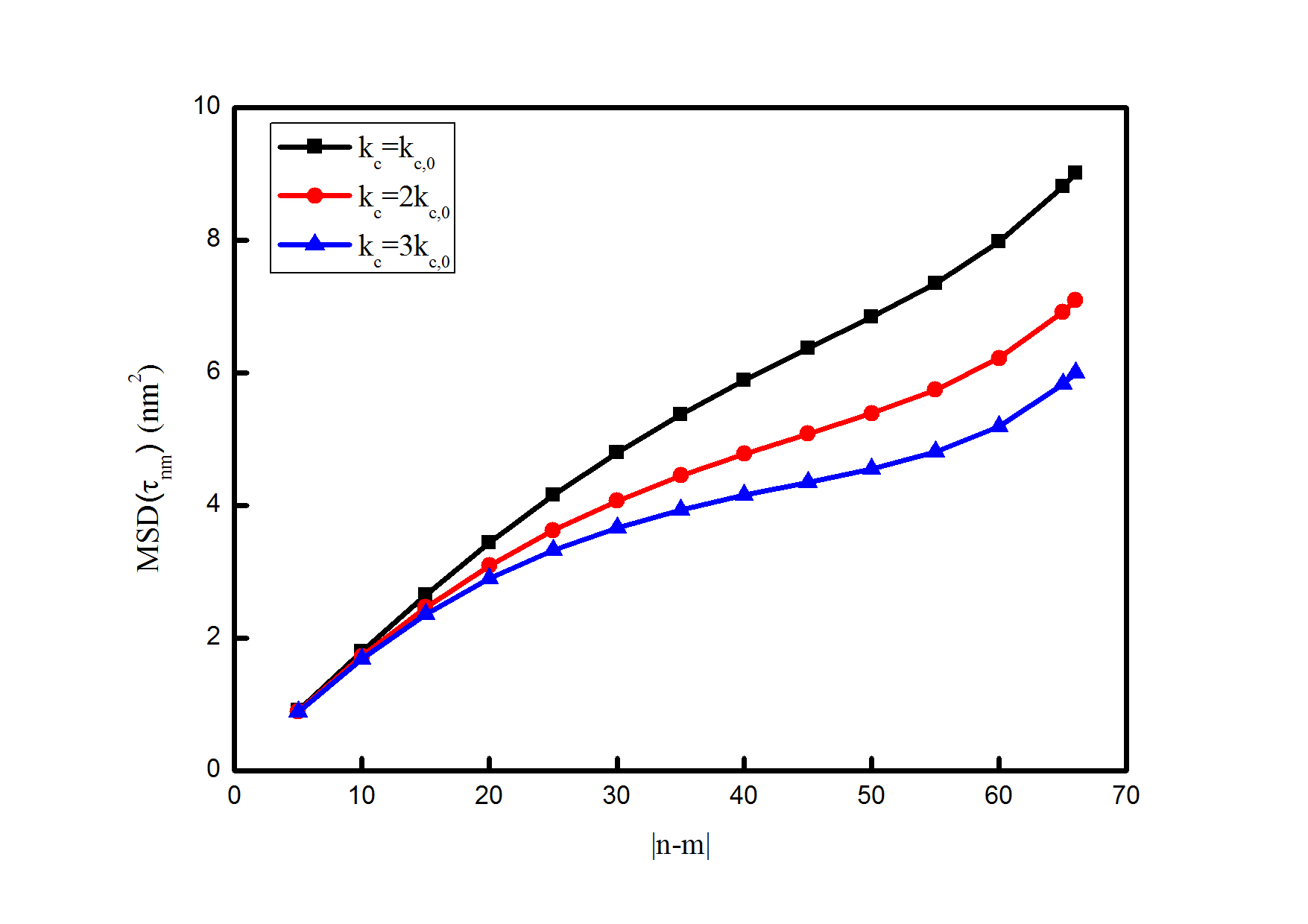}
     \caption{Plot of MSD  at $\tau_{nm}$ vs $|n-m|$ for CRIF. See text for the values of the parameters used.}
     \label{fig:ff}
\end{figure}

Next we plot $\tau_{nm}$ against $|n-m|$, the results obtained here are very fascinating. In case of CRIF, as shown in Fig. (\ref{fig:ee}), when the value of $k_c$ is small $\tau_{nm}$ first increases with increasing $|n-m|$, until it reaches a hump, from where it starts declining. This happens due to two opposing factors. First one is MSD of the distance between the $n^{th}$ and $m^{th}$ monomer, for which an analytical expression is derived in the form of Eq. (\ref{eq:msd}). MSD for every pair of monomers at $t=\tau_{nm}$ is shown in Fig. (\ref{fig:ff}), which clearly shows as the distance between two monomers become greater, the MSD is also higher which results in faster correlation loss of $R_{nm}$. If correlation is decaying fast $\tau_{nm}$ will consequently become smaller. Another factor influencing reconfiguration is the distance between two monomers $n$ and $m$ along the chain. This can be seen from the normal mode description, $R_{nm}$ is a sum of normal modes and as $|n-m|$ decreases, the contribution of higher normal modes  will be greater which relaxes much faster. Thus if $|n-m|$ is small $\tau_{nm}$ should be small. Because of these two competing effects the reconfiguration time passes through the maxima \cite{chakrabarti2014}.

The trend of $\tau_{nm}$ vs $|n-m|$ is quite different at higher degrees of compactness as shown in Fig. (\ref{fig:ee}). Again the plausible reason behind this might be the competing effect of the two factors, MSD and relaxation times of the normal modes describing the part of polymer between the $n^{th}$ and $m^{th}$ monomers. When the value of $k_c$ becomes higher and the polymer becomes more compact. If one looks very carefully into the expression for $\tau_p^{CR(Z)IF}=\frac{\zeta_p^{R(Z)}}{k_p^{CR(Z)IF}}+\frac{\zeta_{int}}{k+k_c N^2/p^2 \pi^2}$, it seems with $k_c$ being independent of $\zeta_{int}$ relaxation should be faster in presence of $k_c$. But $\zeta_{int}$ also becomes large with increasing $k_c$ and as told earlier $\tau_{int}^{CR(Z)IF}\simeq\frac{\zeta_{int}p^2\pi^2}{k_cN^2}$ is greater for higher modes in comparison to the lower ones, which is shown in Fig. (\ref{fig:bb}). In a nutshell slowing down of relaxation time is more for higher normal modes compared to the lower ones in presence of higher $k_c$ . So, at higher $k_c$, compared to lower value, relaxation is more slower when the distance between $n^{th}$ and $m^{th}$ monomers is small, compared to the situation when distance between $n$ and $m$ is longer.  MSD is reduced as well which can be seen in Fig. (\ref{fig:ff}) which is an obvious outcome of compact state of polymer. Because, in this new model beside the polymer being more compact, the movement of each monomer is very restricted. But it is clearly observed from Fig. (\ref{fig:ff}), the decrease of MSD is not consistent. With increasing distance between $n$ and $m$, MSD is more controlled. That is why when $|n-m|$ is small, effect of MSD become more prominent and $\tau_{nm}$ decreases with increasing $|n-m|$ and for higher $|n-m|$ the dynamics of the polymer between $n$ and $m$ is controlled by the the normal modes instead of MSD. As a consequence reconfiguration time increases with increasing $|n-m|$.

\begin{figure}
  \centering
    \includegraphics[width=0.8\textwidth]{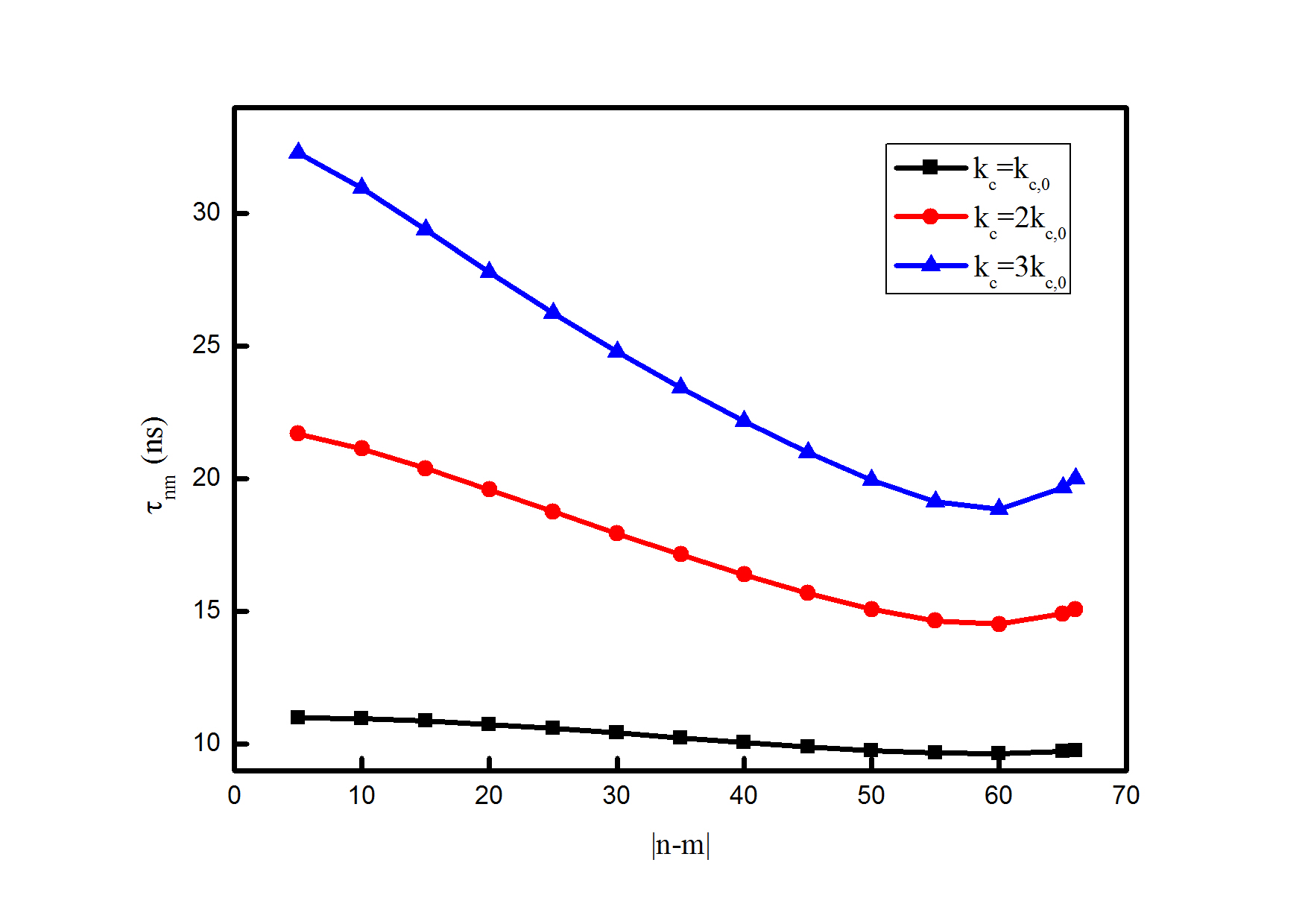}
    \caption{Plot of Reconfiguration time vs $|n-m|$ for CZIF. See text for the values of the parameters used.}
     \label{fig:gg}
\end{figure}

When similar calculations are done for Zimm model, as can be seen in Fig. (\ref{fig:gg}) even for small $k_c$ reconfiguration time decreases with increasing distance between two monomers similar to the second trend discussed for CRIF with higher value of $k_c$. The relaxation time of the normal modes describing Zimm chain are usually faster than that of Rouse \cite{chakrabartiphysica1}. That is why, at even smaller value of $k_c$ the second trend is be observed. This trend is very similar to the experimental results found by Nettels \textit{et al} \cite{schuler2008} where they estimated reconfiguration time as a function of separating distance between two groups from FRET. If the value of $k_c$ is considered even smaller, the first trend observed for CRIF is recovered again which is not shown here.

\begin{figure}
  \centering
    \includegraphics[width=0.8\textwidth]{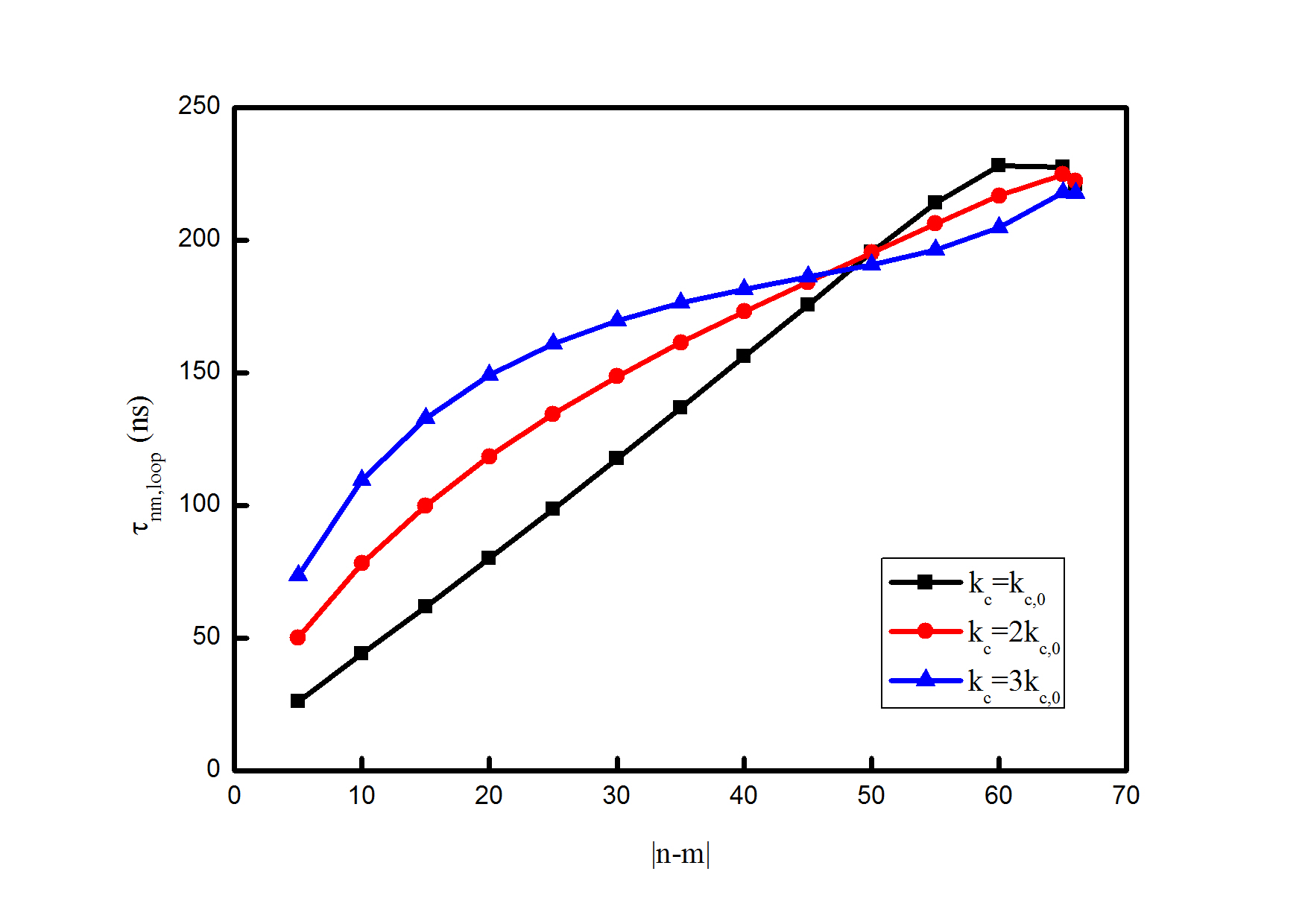}
     \caption{Plot of Looping time vs $|n-m|$ for CRIF. See text for the values of the parameters used.}
     \label{fig:ii}
\end{figure}

\begin{figure}
  \centering
    \includegraphics[width=0.8\textwidth]{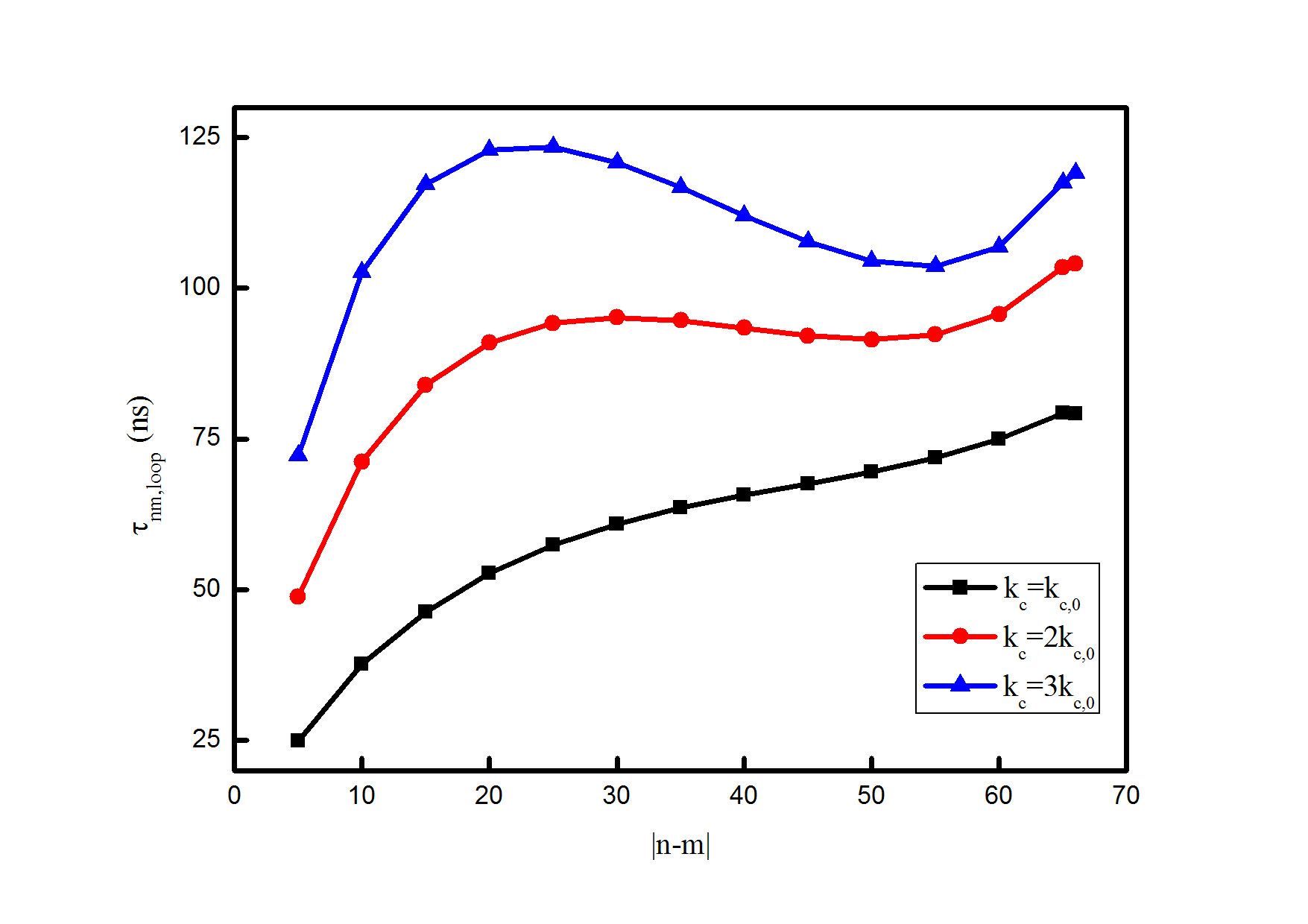}
     \caption{Plot of Looping time vs $|n-m|$ for CZIF. See text for the values of the parameters used.}
     \label{fig:jj}
\end{figure}

Looping time is estimated following the Wilemski Fixman (WF) approach, where a sink is placed between the two monomers which will eventually form a loop and has the closed form generalized expression given in Eq. (\ref{eq:looping-time-mn}). Fig. (\ref{fig:ii}) and Fig. (\ref{fig:jj}) are showing the dependence of $\tau_{nm,loop}$ on $|n-m|$ for CRIF and CZIF respectively. Variations of $\tau_{nm,loop}$ vs $|n-m|$ is presumably due to two competing factors as in the case of $\tau_{nm}$.

\begin{figure}
  \centering
    \includegraphics[width=0.8\textwidth]{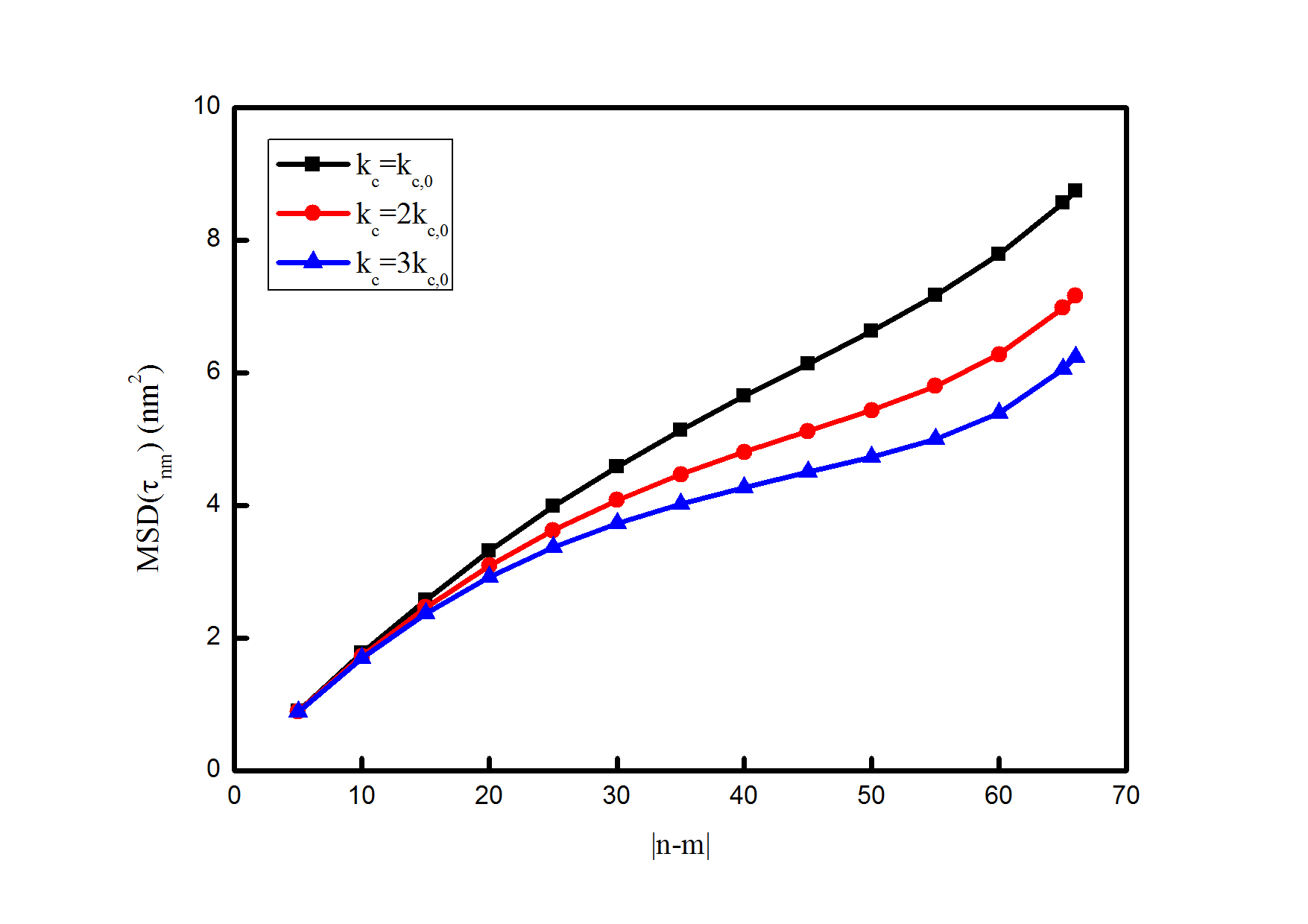}
     \caption{Plot of MSD at $\tau_{nm}$ vs $|n-m|$ for CZIF. See text for the values of the parameters used.}
      \label{fig:hh}
\end{figure}

\section{Conclusion}

The work addresses the looping dynamics of single polymer within WF approximation including the effect of internal friction at various degrees of compactness. Looping and reconfiguration times are calculated using the recently proposed model compacted Rouse with internal friction (CRIF). Same set of calculations are carried out with compacted Zimm with internal friction (CZIF) where hydrodynamic interactions are accounted for. The strength of our method lies in its simplicity. We show that without carrying out expensive simulations even a simple model like ours confirm experimental results. It can account for different intercepts of reconfiguration time ($\tau_{nm}$) vs solvent viscosity ($\eta$) at different denaturant concentrations.  Presently our model does not have a fluctuation-dissipation theorem for the internal friction \cite{schieber2}. Currently we are working towards a model having a fluctuation-dissipation theorem for the internal friction as well. Another aspect is the crowding on looping \cite{metzler2015}. It is expected as crowding increases internal friction should increase. This will be an interesting future problem to investigate.

\section{ACKNOWLEDGEMENT}

N. S. and R. C. thank IRCC IIT Bombay for funding (Project Code: 12IRCCSG046). R. C also thanks DST and CSIR for funding.

\end{document}